\documentclass{article}
\usepackage{amsmath,amssymb,enumerate,bbm,calc,capt-of,ifthen}
\usepackage{graphicx}
\usepackage{epsfig}

\newtheorem{cntr}{do not use}

\newtheorem{definition}[cntr]{Definition}
\newtheorem{assumption}{Assumption}

\newtheorem{varremark}[cntr]{Remark}
\newenvironment{remark}{\begin{varremark}\em}{\em\end{varremark}}
\newcommand{\nin}{\not\in}

\newtheorem{varnote}{Note}

\newtheorem{proposition}[cntr]{Proposition}

\newenvironment{proof}{
  \noindent\textbf{Proof.}\ }{\hspace*{\fill}
  \begin{math}\Box\end{math}\medskip}

\newenvironment{proof*}[1]{
  \noindent\textbf{#1\ }}{\hspace*{\fill}
  \begin{math}\Box\end{math}\medskip}

\newtheorem{lemma}[cntr]{Lemma}

\newtheorem{theorem}[cntr]{Theorem}

\newtheorem{algo}{Algorithm}

\numberwithin{cntr}{section}
\numberwithin{equation}{section}

\newcommand{\comment}[1]{}
\newcommand{\abs}[1]{\left| #1 \right|}
\newcommand{\RR}[0]{{\mathbb{R}}}

\newcommand{\vx}[0]{{\vec{x}}}
\newcommand{\vp}[0]{{\vec{p}}}
\newcommand{\vq}[0]{{\vec{q}}}
\newcommand{\vr}[0]{{\vec{r}}}
\newcommand{\vm}[0]{{\vec{m}}}

\newcommand{\vv}[0]{{\vec{v}}}
\newcommand{\va}[0]{{\vec{a}}}
\newcommand{\vb}[0]{{\vec{b}}}
\newcommand{\vg}[0]{{\vec{g}}}

\newcommand{\Oto}[1]{{0 \ldots #1-1}}
\newcommand{\OtoN}{{0 \ldots N-1}}

\newcommand{\pointData}{{ \{ \vp_{i} \}_{i=\OtoN} }}
\newcommand{\tanData}{{ \{ \vm_{i} \}_{i=\OtoN} }}
\newcommand{\allData}{{ \{ \vp_{i}, \vm_{i} \}_{i=\OtoN} }}

\newcommand{\curveSet}{{ \{ \gamma_i(t) \}_{\Oto{M}}}}
\newcommand{\poly}{{\Gamma}}

\newcommand{\ball}[2]{ { B_{#1}(#2) } }
\newcommand{\allowed}[2]{ { A_{#1}(#2) } }

\newcommand{\curvemax}{{\kappa_{m}}}
\newcommand{\curvemaxi}{{\curvemax^{-1}}}

\newcommand{\curvesep}{{\delta}}

\newcommand{\pointNoise}{{\zeta}}
\newcommand{\tanNoise}{{\xi}}

\newcommand{\nallowed}[2]{ { A^{\pointNoise, \tanNoise}_{#1}(#2) } }

\newcommand{\densitymax}{{\rho_{m}}}

\begin{document}

\title{Reconstructing Curves from Points and Tangents}

\author{L. Greengard and C. Stucchio}

\maketitle

\begin{abstract}
Reconstructing a finite set of curves from an unordered set of sample points
is a well studied topic. There has been less
effort that considers how much better the reconstruction can be if
\emph{tangential} information is given as well.
We show that if curves are separated from each other by a
distance $\curvesep$, then the sampling rate need only be $O(\sqrt{\curvesep})$
for error-free reconstruction.
For the case of point data alone, $O(\curvesep)$ sampling is required.
\end{abstract}

\section{Introduction}

In this paper, we consider the problem of reconstructing a
$C^{1}$ \emph{figure} -- that is, a family of curves $\curveSet$ from a finite
set of data. More precisely, we assume we are given
an unorganized set of points $\pointData$, as well as \emph{unit} tangents to the points $\tanData$. Note that the tangents have no particular orientation; making the change $\vm_{i} \rightarrow -\vm_{i}$ destroys no information.

\begin{definition}
  \label{def:polygonalization}
  A polygonalization of a figure $\curveSet$ is a planar graph
$\Gamma = (V,E)$ with the property that each vertex $p \in V$ is a point on some $\gamma_{i}(t)$, and each edge connects points which are adjacent samples of some curve $\gamma_{i}$.
\end{definition}

\begin{figure}
\setlength{\unitlength}{0.240900pt}
\ifx\plotpoint\undefined\newsavebox{\plotpoint}\fi
\sbox{\plotpoint}{\rule[-0.200pt]{0.400pt}{0.400pt}}%
\includegraphics[scale=0.5]{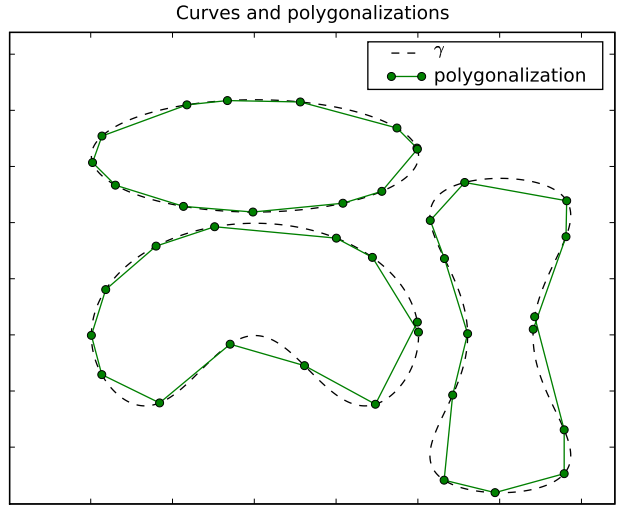}
\caption{A figure and it's polygonalization, c.f. Definition \ref{def:polygonalization}. }
\label{fig:polygonalization}
\end{figure}

Our goal here is to construct an algorithm which reconstructs the
polygonalization of a figure from the data defined above.
An example of a polygonalization is given in Figure \ref{fig:polygonalization}.

The topic of reconstructing figures solely from point data $\pointData$ has been the subject of considerable attention \cite{amenta98crust,amenta98new,dey99curve,hoppe92surface,amenta02simple, dey01reconstructing, edelsbrunner}. This is actually a more difficult problem, and only weaker results are possible. The main difficulty is the following; if the distance between two separate curves $\gamma_{i}$ and $\gamma_{j}$ is smaller than the sample spacing, then it is difficult to determine which points are associated to which curve. Thus, sample spacing must be $O(\curvesep)$, with $\curvesep$ the distance between different curves.

Tangential information makes this task easier; in essence, if two points are nearby (say $\vp_{1}$ and $\vp_{2}$), but $\pm \vm_{1}$ does not point (roughly) in the direction $\vp_{2}-\vp_{1}$, then $\vp_{1}$ and $\vp_{2}$ should not be connected. This fact allows us to reduce the sample spacing to $O(\curvesep^{1/2})$, rather than $O(\curvesep)$. This is to be expected by analogy to interpolation; knowledge of a function and its derivatives yields quadratic accuracy.

We should mention at this point related work on \emph{Surfels} (short for \emph{Surface Elements}). A surfel is a point, together with information characterizing the tangent plane to a surface at that point (and perhaps other information such as texture). They have become somewhat popular in computer graphics recently, mainly for rendering objects characterized by point clouds
\cite{882320,1103907,598521,1018057,344936,383300}.

In this work, we present an algorithm which allows us to reconstruct a curve from $\allData$. We make two assumptions, under which the algorithm is provably correct.

\begin{assumption}
  \label{ass:curvature}
  We assume each curve $\gamma_{i}(t) = (x_i(t),y_i(t))$ has bounded curvature:
  \begin{equation}
    \label{eq:curvatureAssumption}
    \forall i = \Oto{M}, ~ \frac{
      \abs{x_{i}'(t) y_{i}''(t) - y_{i}'(t) x_{i}''(t)}
    } {
      (x_{i}'(t)^{2}+y_{i}'(t)^{2})^{3/2}
    } \leq \curvemax
  \end{equation}
\end{assumption}

This assumption is necessary to prevent the curves from oscillating too much between samples.

\begin{assumption}
  \label{ass:separation}
  We assume the curves $\gamma_{i}$ and $\gamma_{j}$ are uniformly separated from each other, i.e.:
  \begin{subequations}
    \begin{equation}
      \label{eq:separationAssumption}
      \inf_{t,t'} \abs{ \gamma_{i}(t) - \gamma_{j}(t')} \geq \curvesep \textrm{~for~} i \neq j
    \end{equation}
    We also assume that different areas of the same curve are separated
    from each other:
    \begin{equation}
      \label{eq:separationAssumptionSameCurve}
      \inf_{\abs{t-t'} > \curvemaxi\pi/2 } \abs{ \gamma_{i}(t) - \gamma_{i}(t')} \geq \curvesep
    \end{equation}
  \end{subequations}
  (assuming the curve $\gamma_{i}(t)$ proceeds with unit speed).
\end{assumption}

\begin{figure}
\setlength{\unitlength}{0.240900pt}
\ifx\plotpoint\undefined\newsavebox{\plotpoint}\fi
\sbox{\plotpoint}{\rule[-0.200pt]{0.400pt}{0.400pt}}%
\includegraphics[scale=0.5]{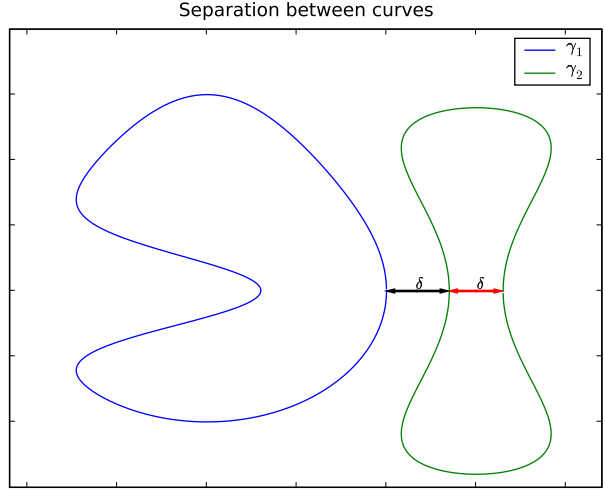}

\caption{An illustration of Assumption \ref{ass:separation}. The black arrow illustrates the condition \eqref{eq:separationAssumption}, while the red arrow illustrates the condition \eqref{eq:separationAssumptionSameCurve}.}
\label{fig:separationBetweenCurves}
\end{figure}

These assumptions ensure that two distinct curves do not come too close
together (\ref{ass:curvature})  and that separate regions of
the same curve do not come
arbitrarily close (\ref{ass:separation}).
This is illustrated in Figure \ref{fig:separationBetweenCurves}.

\section{The Reconstruction Algorithm}

Before we begin, we require some notation.

\begin{definition}
  \label{def:perp}
  For a vector $\vv$, let $\vv^{\perp}$ denote the vector $\vv$
rotated clockwise by an angle $\pi/2$.
\end{definition}

\begin{definition}
  \label{def:metric}
  Let $d(\vp,\vq)$ denote the usual Euclidean metric, $d(\vp,\vq) = \abs{\vp - \vq}$. Let $d_{\vm}(\vp,\vq)$ denote the distance in the $\vm$ direction between $\vp$ and $\vq$, i.e. $d_{\vm}(\vp,\vq) = \abs{ (\vp - \vq) \cdot \vm}$.
\end{definition}

\begin{definition}
  For a point $\vp$ and a curve $\gamma_{i}(t)$, we say that $\vp \in \gamma_{i}(t)$ if $\exists t$ such that $\gamma_{i}(t)=\vp$.
\end{definition}

\subsection{The Forbidden Zone}

Before explaining the algorithm which constructs the polygonalization of
a figure (the set of curves $\curveSet$) from discrete data $\allData$, we
prove a basic lemma which forms the foundation of our method.
We assume for the remainder of this section that the figure
satisfies Assumption 1.

\begin{figure}
\setlength{\unitlength}{0.240900pt}
\ifx\plotpoint\undefined\newsavebox{\plotpoint}\fi
\sbox{\plotpoint}{\rule[-0.200pt]{0.400pt}{0.400pt}}%
\includegraphics[scale=0.5]{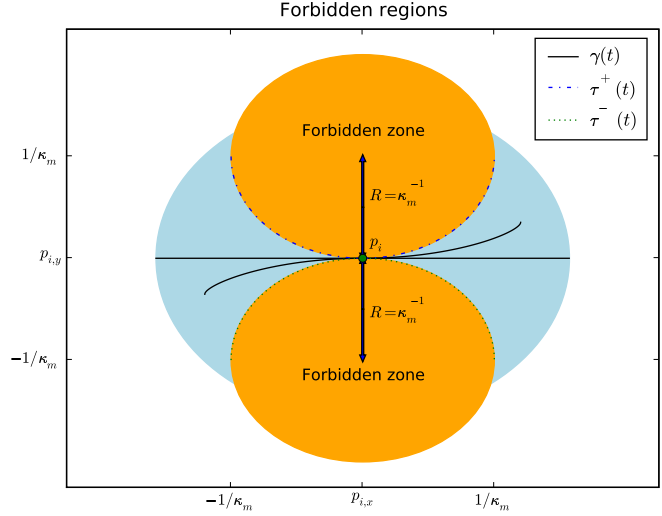}
\caption{The forbidden zones, as described in Lemma \ref{lem:forbiddenZone}.
The orange (darker region) is the forbidden zone, and the blue (lighter region)
is the set of points a distance $\pi \curvemaxi/2$ away from $p_{i}$.}
\label{fig:forbiddenZone}
\end{figure}

\begin{definition}
  For a point $\vp_{i}$, we refer to the set
$\cup_{\pm} \ball{\curvemaxi}{\vp_{i} \pm \vm_{i}^{\perp} \curvemaxi}$
as its \emph{forbidden zone},
illustrated in Fig. \ref{fig:forbiddenZone}.
Here, $\ball{r}{\vp}$ is the usual ball of radius $r$ about $\vp$.
\end{definition}

\begin{lemma}
\label{lem:forbiddenZone}
For every $i \neq j$, if $\vp_{j}$ is in the forbidden zone of $\vp_i$,
then $(\vp_{i},\vp_{j})$ is not an edge in $\poly$ assuming that the
sample spacing is less than $\curvemaxi \pi/2$.
\end{lemma}

\begin{proof}
Suppose for simplicity that $\vp_{i}=(0,0)$ and $\vm_{i}=(1,0)$. Now, consider a line $\tau(t)$ of maximal curvature. The curve of maximal curvature, with $\tau_{y}'(t) > 0$ and proceeding at speed $\curvemaxi$ is $\tau^{+}(t)=(\curvemaxi \sin(t), \curvemaxi (1-\cos(t)))$, while the curve with $\tau_{y}'(t) < 0$ is $\tau^{-}(t)=(\curvemaxi \sin(t), \curvemaxi (\cos(t)-1))$.

By assumption 1, the curve $\gamma(t)$ containing $\vp_i$
must lie between these curves (the near boundaries of the forbidden zone
in Fig \ref{lem:forbiddenZone}). Thus, it is confined to the blue (lighter)
region while its arc length is less than $\curvemaxi \pi/2$.
If $\vp_{j}$ is in the forbidden zone and
$\gamma(t)$ connects $\vp_{i}$ to $\vp_{j}$, then it must do so after travelling a distance greater than $\curvemaxi \pi/2$.
\end{proof}

\begin{figure}
\setlength{\unitlength}{0.240900pt}
\ifx\plotpoint\undefined\newsavebox{\plotpoint}\fi
\sbox{\plotpoint}{\rule[-0.200pt]{0.400pt}{0.400pt}}%
\includegraphics[scale=0.5]{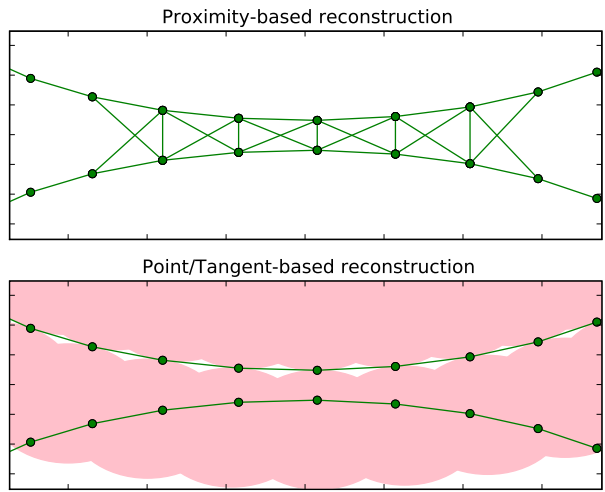}

\caption{A naive proximity-based reconstruction algorithm (shown),
or even a $\beta$-crust type algorithm, will introduce edges
between different curves. Knowledge of the forbidden zone allows us to
remove such edges.}
\label{fig:proximityVsTangentBased}
\end{figure}

In short, the extra information provided by the tangents
allows us to exclude edges from the polygonalization if they point too far away
from the tangent, resulting in higher fidelity (c.f. Fig. \ref{fig:proximityVsTangentBased}).

\begin{definition}
  \label{def:AllowedRegion}
For a point $\vp$, we define the \emph{allowed zone} or
\emph{allowed region} $\allowed{\epsilon}{\vp}$ by
  \begin{equation}
    \label{eq:allowedRegion}
    \allowed{\epsilon}{\vp}=\ball{\epsilon}{\vp_{i}} \setminus \left[ \cup_{\pm} \ball{\curvemaxi}{\vp_{i} \pm \vm_{i}^{\perp} \curvemaxi} \right]
  \end{equation}
  That is, $\allowed{\epsilon}{\vp}$ is the ball of radius $\epsilon$ about $p$ excluding the forbidden zone.
\end{definition}

Clearly, any edge in the polygonalization starting at $\vp$, with length shorter than $\epsilon$, must connect to another point $\vq \in \allowed{\epsilon}{\vp}$.  We are now ready to describe the polygonalization algorithm.

\vspace{.2in}

\hrule
\begin{algo}
  \label{algo:polygonalization}
  \begin{center} {\bf (Noise-Free Polygonalization)}
  \end{center}

\vspace{.1in}

\hrule

\vspace{.2in}

\noindent { \bf Input: }
[ We assume we are given
the dataset $\allData$,
the maximal curvature $\curvemax$, and
a parameter $\epsilon$ satisfying both
$\epsilon \curvemax < 1/\sqrt{2}$ and $2 \curvemax \epsilon^{2} < \curvesep$.
We assume that adjacent points on a given curve
are less than a distance $\epsilon$ apart, i.e. the curve is
$\epsilon$-sampled. ]

\vspace{.1in}

  \begin{enumerate}
  \item Compute the graph $G = (\pointData, E)$ with edge set:
    \begin{equation*}
      E = \{ (\vp_{i},\vp_{j}) : \vp_{i} \in \allowed{\epsilon}{\vp_{j}} \textrm{~and~} \vp_{j} \in \allowed{\epsilon}{\vp_{i}}\}
    \end{equation*}
  \item For each vertex $\vp_{i} \in \pointData$:
    \begin{enumerate}[a.]
    \item Compute the set of vertices
      \begin{equation*}
        R^{\pm}_{i} = \{ \vp_{j} : (\vp_{i}, \vp_{j}) \in E \textrm{~and~} \pm (\vp_{j}-\vp_{i}) \cdot \vm_{i} > 0 \}
      \end{equation*}
    \item Find the nearest tangential neighbors, i.e.
      \begin{equation*}
        \vr^{\pm}_{i} = \textrm{argmin}_{\vq \in R^{\pm}_{i}} d_{\vm_{i}}(\vq, \vp_{i})
      \end{equation*}
    \end{enumerate}
  \item Output the graph $\Gamma = ( \pointData, E')$ with
    \begin{equation*}
      E' = \{ (\vp_{i}, \vr^{+}_{i}) \} \cup \{ (\vp_{i}, \vr^{-}_{i}) \}
    \end{equation*}
    This graph is the polygonalization of $\curveSet$.
  \end{enumerate}

\end{algo}

\begin{remark}
  As presented, the complexity of Algorithm \ref{algo:polygonalization} is $O(N^{2})$, due to both step 1 and step 2. (Step 2 can be slow if $O(N)$ points are within the allowed region of some particular point). The complexity can be reduced to $O(N \log N)$ using quadtrees if we assume a minimal sampling rate (see Appendix \ref{sec:quadTreeSection}).
\end{remark}

The following theorem guarantees the correctness of
Algorithm \ref{algo:polygonalization}. Its proof is presented
in the next section.

\begin{theorem}
  \label{thm:proofOfAlgo}
  Suppose that:
  \begin{subequations}
    \label{eq:separationCondition}
    \begin{equation}
      \label{eq:constraintOnSeparationSampling}
      \curvesep > 2\curvemax \epsilon^{2}
    \end{equation}
    where $\curvesep$ is as in Assumption \ref{ass:separation} and also
    \begin{equation}
      \label{eq:constraintOnkmaxEpsilon}
      \epsilon < \frac{1}{ \curvemax \sqrt{2}}
    \end{equation}
  \end{subequations}
  Suppose also that the distance between adjacent samples in the polygonalization is bounded by $\epsilon$, i.e. the curve is $\epsilon$-sampled. Then graph $\Gamma$ returned by Algorithm \ref{algo:polygonalization} is the polygonalization of $\curveSet$.
\end{theorem}

\subsection{Proof of Theorem \ref{thm:proofOfAlgo}}

\begin{lemma}
  \label{lem:separationAllowedRegions}
  Suppose $i \neq j$ and that Assumption \ref{ass:separation} holds.
Then for all $t, t'$, if \eqref{eq:separationCondition} holds, then
  \begin{equation}
    \label{eq:connectionsBetweenDifferentCurvesNotAllowed}
    \gamma_{j}(t')  \nin  \allowed{\epsilon}{\gamma_{i}(t)}.
  \end{equation}
  Similarly, if $i=j$ and $\abs{t-t'} \geq \curvemax^{-1} \pi/2$, then \eqref{eq:connectionsBetweenDifferentCurvesNotAllowed} holds.
\end{lemma}
\begin{proof}
  Fix $t$, and define $\vp=\gamma_{i}(t)$ and $\vm=\gamma_{i}'(t) / \abs{\gamma_{i}'(t)}$. Define $L$ to be the line segment $L= \{ \vp+\vm \curvemaxi \sin(\theta) : \theta \in [-\arcsin(\epsilon \curvemax),\arcsin(\epsilon \curvemax)] \}$. The boundaries of $\allowed{\epsilon}{\vp}$ are given by
\begin{equation*}
  \vp + \vm \curvemaxi \sin(\theta) \pm \vm^{\perp} \curvemaxi (1-\cos(\theta)).
\end{equation*}
Now, for any $\vq \in \gamma_{i} $ and $\vq \in \allowed{\epsilon}{\vp}$, the distance between $\vq$ and $L$ is the normal distance to $L$. This distance is bounded by:
\begin{multline}
  \label{eq:1}
  d(\vq,L) \leq
  \sup_{\theta} \curvemaxi \abs{(1-\cos(\theta)) }\\
  \leq
  \sup_{\theta} \curvemaxi 2 \sin^{2}(\theta/2) =
  2\curvemaxi \sin^{2}( \arcsin(\epsilon \curvemax)/2)
\end{multline}
The intermediate value theorem implies $\arcsin( x) \leq \arcsin'(\zeta) x=(1-\zeta^{2})^{-1/2} x$ for some $\zeta \in [0,x]$; since $\epsilon \curvemax < 2^{-1/2}$ (by \eqref{eq:constraintOnkmaxEpsilon}), we find that:
\begin{equation*}
  \arcsin(\epsilon \curvemax) \leq (1-(2^{-1/2})^{2})^{-1/2} \curvemax \epsilon = \sqrt{2} \curvemax \epsilon
\end{equation*}
Substituting this into \eqref{eq:1} yields:
\begin{equation}
  \label{eq:4}
  d(\vq,L) \leq
  2 \curvemaxi \sin^{2}( \sqrt{2} \curvemax \epsilon/2) \leq \curvemax \epsilon^{2}
\end{equation}
Thus, the \emph{normal} distance between any point in $\allowed{\epsilon}{\vp}$ and $L$ is $O(\curvemax \epsilon^{2})$.

If $\gamma_{j}(t') \nin L+\vm^{\perp} \RR$,
then clearly $\gamma_{j}(t') \nin \allowed{\epsilon}{\gamma_{i}(t)}$
so we assume $\gamma_{j}(t') \in L+\vm^{\perp} \RR$. In this case,
$\gamma_{j}(t') = \vp + \vm \curvemaxi \sin(\theta_{0}) + \vm^\perp z_j$
for some $\theta_0 \in
[-\arcsin(\epsilon \curvemax),\arcsin(\epsilon \curvemax)]$
and $z_j \in \RR$.
Thus, $|z_j| = d_{\vm^{\perp}}(\gamma_{j}(t'), L)$,
the normal distance to $L$. By construction, there is a unique
value $t_i'$ such that
$\gamma_{i}(t_i') = \vp + \vm \curvemaxi \sin(\theta_{0}) + \vm^\perp z_i$.
$|z_i|$ then equals $d_{\vm^{\perp}}(\gamma_{i}(t), L)$.
By the second triangle inequality,
\begin{equation*}
  d_{\vm^{\perp}}(\gamma_{j}(t'), L)  = |z_j|
  \geq \abs{|z_j - z_i| - |z_i|}
  \geq \delta - \curvemax \epsilon^{2} > \curvemax \epsilon^{2}
\end{equation*}
But this implies that $d(\gamma_{j}(t'), L) \geq d_{\vm^{\perp}}(\gamma_{j}(t'), L) \geq \curvemax \epsilon^{2}$, and thus $\gamma_{j}(t') \nin \allowed{\epsilon}{\vp}$.

The proof when $i=j$ is identical.
\end{proof}

This result shows that the graph $G$, computed in Step 1 of Algorithm \ref{algo:polygonalization}, separates different $\gamma_{i}$ and $\gamma_{j}$ from each other, as well as different parts of the same curve. Thus, after Step 1, we are left with a graph $G$ having edges only between points $\vp_{i}$ and $\vp_{j}$ which are on the same curve $\gamma_{k}$, and which are separated along $\gamma_{k}$ by an arc length no more than $\curvemaxi \pi/2$.

We now show that $G$ is a superset of the polygonalization $\Gamma$.

\begin{proposition}
  \label{prop:polyIncludesNeighboringPoints}
  Suppose the point data $\pointData$ is $\epsilon$-sampled, i.e. if two points $\vp_{i}$ and $\vp_{j}$ are adjacent on the curve $\gamma_{k}$, then the \emph{arc length} between $\vp_{i}$ and $\vp_{j}$ is bounded by $\epsilon$. Then $G$ contains the polygonalization of $\curveSet$.
\end{proposition}
\begin{proof}
  If the distance between adjacent points $\vp_{i}$ and $\vp_{j}$ is at most $\epsilon$, then $\vp_{j} \in \ball{\epsilon}{\vp_{i}}$. Since the segment of $\gamma_{k}$ between $\vp_{i}$ and $\vp_{j}$ has arc length less than $\epsilon$, $\vp_{j}$ is not in the forbidden zone of $\vp_{i}$ (by the same argument as in Lemma \ref{lem:forbiddenZone}. Thus, $\vp_{j} \in \allowed{\epsilon}{\vp_{i}}$ (and vice versa), and $(\vp_{i},\vp_{j})$ is an edge in $G$.
\end{proof}

We have now shown that $G$ separates distinct curves, and that $G$ contains the polygonalization $\Gamma$ of $\curveSet$. It remains to show that $G = \Gamma$.

\begin{lemma}
  \label{lem:localGraphParameterization}
  A curve $\gamma_{i}(t)$ satisfying \eqref{eq:curvatureAssumption} admits the local parameterization
  \begin{equation}
    \label{eq:localGraphParameterization}
    \gamma_{i}(t) = \gamma(t_{0}) + (t-t_{0})\gamma'(t_{0}) + w(t) \gamma'^{\perp}(t_{0})
  \end{equation}
  where $w'(t_{0})=0$. The parameterization is valid for $\abs{t-t_{0}} < \curvemaxi$. In particular, $w(t) < f^{-1}(\curvemax t) $ where $f(z)=z / \sqrt{1+z^{2}}$.
\end{lemma}

\begin{proof}
  Taylor's theorem shows the parameterization to be valid on an arbitrarily small ball. All we need to do is show that this parameterization is valid on a region of size $\curvemaxi$.

  The parameterization breaks down when $w'(t)$ blows up, so we need to show that this does not happen before $t=\epsilon$. Plugging this parameterization into the curvature bound \eqref{eq:curvatureAssumption} yields:
  \begin{equation*}
    \frac{ \abs{w''(t)} }{(1+w'(t)^{2})^{3/2}} \leq \curvemax
  \end{equation*}
  Assuming $w''(t)$ is positive, this is a first order nonlinear differential inequality for $w'(t)$. We can integrate both sides (using the hyperbolic trigonometric substitution $w(t)=\sinh(\theta)$ for the left side) to obtain:
  \begin{equation}
    \label{eq:2}
    \frac{w'(t)}{\sqrt{1+w'(t)^{2}}} \leq \curvemax t \, .
  \end{equation}
  With $f(z)$ defined as in the statement, then $f^{-1}(z)$ is singular only at $z=\pm 1$, and is regular before that. Solving \eqref{eq:2} for $w'(t)$ shows that:
  \begin{equation*}
    w'(t) \leq f^{-1}(\curvemax t)
  \end{equation*}
  implying that $w'(t)$ is finite for $\curvemax t < 1$, or $t < \curvemaxi$.
\end{proof}

\begin{lemma}
  \label{lem:closestTangentPointInAllowedRegionIsCorrect}
  Fix a point $\vp_{i}=\vp \in \pointData$. Choose a tangent vector $\vm_{0}$ and fix an orientation, say $+$. Consider the set of points $\vp_{j}$ such that $(\vp, \vp_{j})$ is an edge in $G$ and $(\vp_{j} - \vp) \cdot \vm_{0} > 0$. Suppose also that $\epsilon$ satisfies \eqref{eq:constraintOnkmaxEpsilon}.
Then, the only edge in the polygonalization of $\gamma$ is the edge for which $(\vp_{j} - \vp) \cdot \vm_{0}$ is minimal.
\end{lemma}

\begin{proof}
  By Lemma \ref{lem:localGraphParameterization}, the curve $\gamma(t)$ can be locally parameterized as a graph near $\vp$, i.e. \eqref{eq:localGraphParameterization}. This is valid up to a distance $\curvemaxi$; by \eqref{eq:constraintOnkmaxEpsilon}, it is valid for all points in the graph $G$ connected to $\vp$.

The adjacent points on the graph are the ones for which $\abs{t-t_{0}}$ is minimal. Note that $\vm_{0} \cdot (\vp_{j} - \vp) = t$ (simply plug in \eqref{eq:localGraphParameterization}); thus, minimizing $\vm_{0} \cdot (\vp_{j} - \vp)$ selects the adjacent point on the graph.
\end{proof}

The minimal edge is the edge $\vr^{+}_{0}$ as computed in Step (2b) of Algorithm \ref{algo:polygonalization}.
Thus, we have shown that the computed graph $G$ is the polygonalization
$\Gamma$ of $\curveSet$.

\begin{figure}
\setlength{\unitlength}{0.240900pt}
\ifx\plotpoint\undefined\newsavebox{\plotpoint}\fi
\sbox{\plotpoint}{\rule[-0.200pt]{0.400pt}{0.400pt}}%
\includegraphics[scale=0.5]{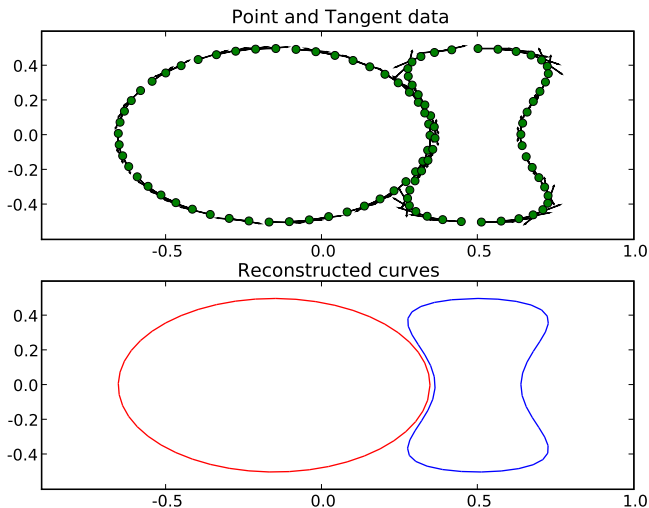}

\caption{Some unordered points/tangents, and the curves reconstructed from them. In this case, $\epsilon=0.065$, $\curvemax=3$ and $\delta=0.015$.}
\label{fig:basicExample}
\end{figure}

\section{Reconstruction in the Presence of Noise}

In practice one rarely has perfect data, so it is important to understand the performance of the approach in the presence of errors.
To that end, we consider the polygonalization problem, but with the point data perturbed by noise smaller than $\pointNoise$ and the tangent data perturbed by noise smaller than $\tanNoise$.
By this we mean the following; to each point $\vp_{i} \in \pointData$, there exists a point $\vp_{i,\ast} = \gamma_{k_{i}}(t_{i})$ such that $\abs{\vp_{i}-\vp_{i,\ast}} \leq \pointNoise$. Similarly, the unit tangent vector $\vm_{i}$ differs from the true tangent $\vm_{i,\ast} = \gamma_{k_{i}}'(t_{i})$ by an angle at most $\tanNoise$. By a polygonalization of the noisy data, we mean that $(\vp_{i},\vp_{j})$ is an edge in the noisy polygonalization if $(\vp_{i,\ast},\vp_{j,\ast})$ is an edge in the noise-free polygonalization. In what follows, $\vp_{j}$ refers to a given (noisy) point, while $\vp_{j,\ast}$ refers to the corresponding true point (and similarly for tangents).

Noise, of course, introduces a lower limit on the features we can resolve. At the very least, the curves must be separated by a distance greater than or equal to $\pointNoise$, to prevent noise from actually moving a sample from one curve to another. In addition, noise in the tangent data introduces uncertainty which forces us to increase the sampling rate; in particular, we require $O(\epsilon \tanNoise + \epsilon^{2}) < \curvesep$.

The main idea in extending Algorithm \ref{algo:polygonalization} to the noisy case is to expand the allowed regions to encompass all possible points and tangents. Of course, this imposes new constraints on the separation between curves.

We also require a \emph{maximal} sampling rate in order to ensure that the order of points on the curve is not affected by noise.
For work in the context of reconstruction using point samples only, see \cite{chengnoise,mdnoise}.

\begin{assumption}
  \label{ass:minSamplingRateNoisy}
  We assume that adjacent points $\vp_{i}$ and $\vp_{j}$ on the curve $\gamma_{k}(t)$ are separated by a distance greater
  than $[(1+2^{3/2})(2 \tanNoise \epsilon + \pointNoise)]$.
\end{assumption}

To compensate for noise, we expand the allowed region to account
for uncertainty concerning the actual point locations.

\begin{definition}
  \label{def:AllowedRegionNoisy}
The \emph{noisy allowed region} $\nallowed{\epsilon}{\vp_i}$
is the union of the allowed regions of all points/tangents
near $(\vp_i, \vm_i)$:
  \begin{equation}
    \label{eq:allowedRegionNoisy}
    \nallowed{\epsilon}{\vp_i}= \bigcup_{
      \substack{
        \abs{\vp -\vp_{i}} < \pointNoise\\
        \arccos(\vm_{i} \cdot \vm) < \tanNoise
      }
    }
    \left(
      \ball{\epsilon}{\vp} \setminus
\left[ \cup_{\pm} \ball{\curvemaxi}{\vp
\pm \vm^{\perp} \curvemaxi} \right]
    \right)
  \end{equation}
\end{definition}

\vspace{.2in}

\hrule
\begin{algo}
  \label{algo:polygonalizationNoisy}
  \begin{center} {\bf (Noisy Polygonalization)}
  \end{center}

\vspace{.1in}

\hrule

\vspace{.2in}

\noindent { \bf Input: }
[ We assume we are given the dataset $\allData$, the maximal curvature
$\curvemax$, the noise amplitudes $\pointNoise, \tanNoise$, and a
parameter $\epsilon$ satisfying both $\epsilon \curvemax < 1/\sqrt{2}$ and
$4 \pointNoise + 2 \epsilon \tanNoise + 2.1 \curvemax \epsilon^{2} < \curvesep$. We assume that adjacent points on a given curve
are less than a distance $\epsilon$ apart, i.e. the curve is
$\epsilon$-sampled. ]

\vspace{.1in}

\begin{enumerate}
  \item Compute the graph $G = (\pointData, E)$ with edge set:
    \begin{equation}
      \label{eq:noisyConditionForCheckingIfEdgeConnectionIsPlausible}
      E = \{ (\vp_{i},\vp_{j}) : \ball{\pointNoise}{\vp_{i}} \cap \nallowed{\epsilon}{\vp_{j}} \neq \emptyset  \textrm{~and~} \ball{\pointNoise}{\vp_{j}} \cap \nallowed{\epsilon}{\vp_{i}} \neq \emptyset \}
    \end{equation}
  \item For each vertex $\vp_{i} \in \pointData$:
    \begin{enumerate}[a.]
    \item Compute the set of vertices
      \begin{equation*}
        R^{\pm}_{i} = \{ \vp_{j} : (\vp_{i}, \vp_{j}) \in E \textrm{~and~} \pm (\vp_{j}-\vp_{i}) \cdot \vm_{i} > 0 \}
      \end{equation*}
    \item Find the nearest tangential neighbors, i.e.
      \begin{equation*}
        \vr^{\pm}_{i} = \textrm{argmin}_{\vq \in R^{\pm}_{i}} d_{\vm_{i}}(\vq, \vp_{i})
      \end{equation*}
    \end{enumerate}
  \item Output the graph $\Gamma = ( \pointData, E')$ with
    \begin{equation*}
      E' = \{ (\vp_{i}, \vr^{+}_{i}) \} \cup \{ (\vp_{i}, \vr^{-}_{i}) \}
    \end{equation*}
    This graph is the polygonalization of $\curveSet$.
  \end{enumerate}

\end{algo}

\begin{figure}
\setlength{\unitlength}{0.240900pt}
\ifx\plotpoint\undefined\newsavebox{\plotpoint}\fi
\sbox{\plotpoint}{\rule[-0.200pt]{0.400pt}{0.400pt}}%
\includegraphics[scale=0.50]{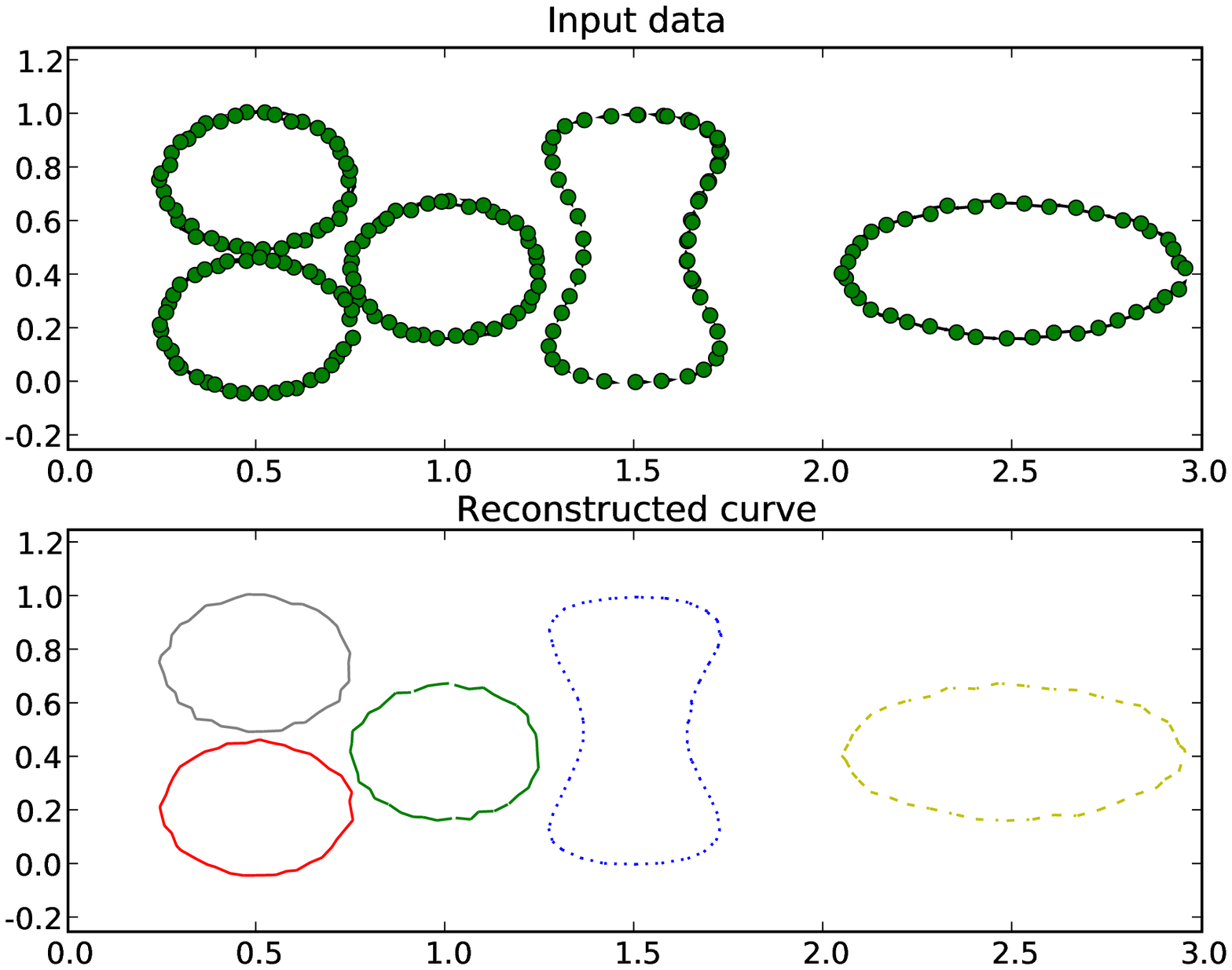}
\caption{Noisy sampled points, and the reconstruction by Algorithm \ref{algo:polygonalizationNoisy}. This example takes $\curvemax=5$, $\epsilon=0.15$, $\pointNoise=\tanNoise=0.01$. }
\label{fig:noisyreconstruction}
\end{figure}

The following theorem
guarantees that Algorithm \ref{algo:polygonalizationNoisy} works.
The proof follows that of Theorem \ref{thm:proofOfAlgo}
and is given in Appendix \ref{sec:proofOfNoisyReconstruction}.
An application is shown in Fig. \ref{fig:noisyreconstruction}.

\begin{theorem}
  \label{thm:noisyReconstruction}
  Suppose that Assumptions \ref{ass:curvature}, \ref{ass:separation} and \ref{ass:minSamplingRateNoisy} hold. Suppose also that
  \begin{subequations}
    \label{eq:noisySeparationConditions}
    \begin{equation}
      \label{eq:noisySeparationDistance}
      \curvesep > 4 \pointNoise + 4 \epsilon \tanNoise + 2.1 \curvemax \epsilon^{2} \, ,
    \end{equation}
    \begin{equation}
      \label{eq:noisyConstraintOnkmaxEpsilon}
      \epsilon < \frac{1}{ \curvemax \sqrt{2}} \, .
    \end{equation}
  \end{subequations}
  Then, Algorithm \ref{algo:polygonalizationNoisy} correctly reconstructs
the figure.
\end{theorem}

\begin{remark}
Consider a point $\vp$, which is a noisy sample from some curve in the figure.
All we can say a priori is that $\vp$ is close to the true
sample $\vp_{\ast}$, i.e. $\vp \in \ball{\pointNoise}{\vp_{\ast}}$.
However, given the knowledge that the polygonalization contains
the edges $(\vq,\vp)$ and $(\vp,\vr)$, we can obtain further information
on $\vp_{\ast}$. Not only does $\vp_{\ast}$ lie in $\ball{\pointNoise}{\vp}$,
but $\vp_{\ast} \in \nallowed{\epsilon}{\vq}$ and
$\vp_{\ast} \in \nallowed{\epsilon}{\vr}$. In short,
\begin{equation}
  \label{eq:noisyFilteringFromAllowedRegions}
  \vp_{\ast} \in \ball{\pointNoise}{\vp} \cap \nallowed{\epsilon}{\vq} \cap  \nallowed{\epsilon}{\vr}
\end{equation}
We can therefore improve our approximation to $\vp_{\ast}$ by minimizing
either the worst case error,
\begin{subequations}
  \begin{equation}
    \vp^{new} = \textrm{argmin}_{\vp}  \sup_{\vx \in A} \abs{\vp - \vx}, ~ A = \ball{\pointNoise}{\vp} \cap \nallowed{\epsilon}{\vq} \cap  \nallowed{\epsilon}{\vr}
  \end{equation}
  or the mean error,
  \begin{equation}
    \vp^{new} = \textrm{argmin}_{\vp}  \int_{A} \abs{\vp - \vx} d\vx
  \end{equation}
\end{subequations}
or some application-dependent functional.
Noise in the tangential data can be similarly reduced.
This is a postprocessing matter after polygonalization,
and we will not expanded further on this idea in the present paper.
\end{remark}

\section{Examples}

\subsection{Extracting Topology from MRI images}

In its simplest version, Magnetic Resonance Imaging (MRI) is
used to obtain the
two-dimensional Fourier transform of the proton density in a
planar cross-section through the patient's body.
That is, if $\rho(x)$ is
is the proton density distribution in the plane $P$, then the MRI device
is able to return the data $\hat{\rho}(k)$ at a selection of
points $k$ in the Fourier transform domain ($k$-space).
The number of sample points available, however, is finite and covers
only the low-frequency range in $k$-space well.
Thus, it is desirable to be able to make use of the limited
information in an optimal fashion.
We are currently exploring methods for MRI based on
exploiting the assumption that
$\rho(x)$ is piecewise smooth (since different tissues have different
densities, and the tissues boundaries tend to be sharp).
Our goal is to carry out reconstruction in three steps.
First, we find the tissue boundaries (the discontinu-
ities). Second, we subtract the influence of the discontonuities from
the measured $k$-space data and third, we reconstruct the remainder which
is now smooth (or smoother). Standard filtered Discrete Fourier Transforms
are easily able to reconstruct the remainder, so the basic
problem is that of reconstructing the edges.

\begin{figure}
\setlength{\unitlength}{0.240900pt}
\ifx\plotpoint\undefined\newsavebox{\plotpoint}\fi
\sbox{\plotpoint}{\rule[-0.200pt]{0.400pt}{0.400pt}}%
\includegraphics[scale=0.5]{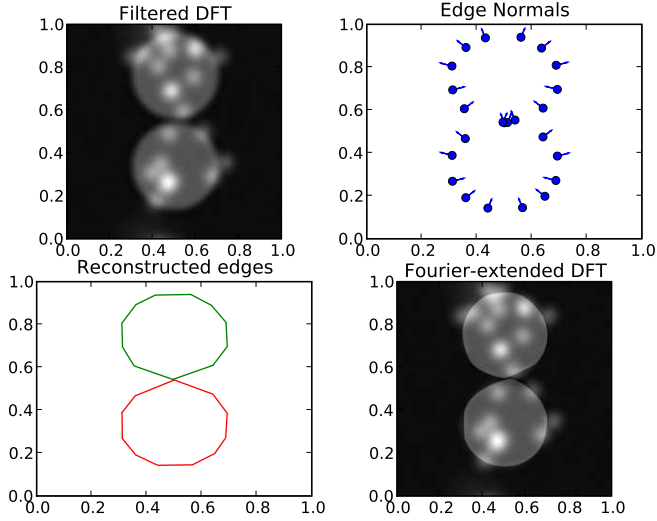}
\caption{A simulated MRI image. The original image was two circles, together with some low frequency ``texture''. The noise level is 5\%.}
\label{fig:mriExample}
\end{figure}

Using directional edge detectors on the $k$-space data, we can extract
a set of point samples from the edges, together with non-oriented normal
directions.  By means of
Algorithm \ref{algo:polygonalizationNoisy}, we can
reconstruct the topology of the edge set and carry out the
procedure sketched out above.
The details of the algorithm are beyond the scope of this article,
and will be reported at a later date,
but Figure \ref{fig:mriExample}
illustrates the idea behind the method. Our work on curve reconstruction was,
in fact, motivated by this application.

\subsection{Figure detection}

\begin{figure}
\setlength{\unitlength}{0.240900pt}
\ifx\plotpoint\undefined\newsavebox{\plotpoint}\fi
\sbox{\plotpoint}{\rule[-0.200pt]{0.400pt}{0.400pt}}%
\includegraphics[scale=0.5]{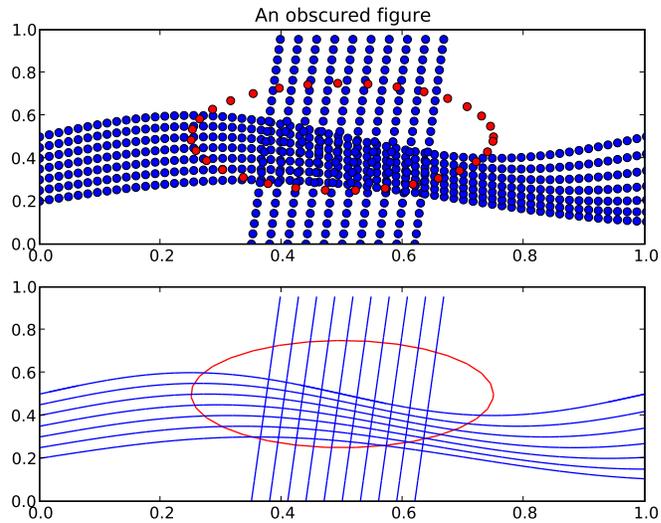}
\caption{A figure which is partially obscured.
Algorithm \ref{algo:polygonalization} correctly computes its
polygonalization, and distinguishes it from the curves in front of it.
To avoid visual clutter, the tangents are not displayed in this figure.}
\label{fig:obscuredExample}
\end{figure}

A natural problem in various computer vision applications is that of
recognizing sampled objects that are partially obscured by a
complex foreground.
As a model of this problem, we constructed an (oval)
figure, and obscured it by covering it with a sequence of curves.
Algorithm \ref{algo:polygonalization} succesfully reconstructs the
figure, as well as properly connecting points on the
horizontally and vertically oriented covering curves.
The result is shown in
Figure \ref{fig:obscuredExample}. Note that the branches are not
connected to the oval (or each other).

\subsection{Filtering spurious points}

The method provided here is relatively robust with regard to the
addition of spurious random data points. This is because spurious data
points are highly unlikely to be connected to any other points in the
polygonalization graph. To see this, note
first that for an incorrect data point to be connected to part of the
polygonalization at all, it would need to be located in
$\allowed{\epsilon}{\vp}$ for some $\vp$.
This is a region of length $O(\epsilon)$ and width $O(\epsilon^{2})$.
There are approximately $L = \sum_{j} \textrm{arclength}(\gamma_{j})$
such points, for a total volume of $\epsilon^{2} L$. Thus, the probability
that a spurious point is in \emph{some} allowed region is roughly
$O(L \epsilon^{2})$.

The second reason is that even if a spurious point is in some allowed region,
it is unlikely to point in the correct direction.
If an erroneous point $\vq$ is inside $\allowed{\epsilon}{\vp}$, it is
still not likely that $\vp \in \allowed{\epsilon}{\vq}$, since
the tangent at $\vq$ must point in the direction of $\vp$
(with error proportional to $\epsilon^{2}$, the angular width of
$\allowed{\epsilon}{\vq}$). Thus, the probability that the tangent at
$\vq$ points towards $\vp$ is $O(\epsilon^{2}/2\pi)$.
Combining these arguments, the probability that any \emph{randomly chosen}
spurious point $\vq$ is connected to any other point in the
polygonalization is $O(L \epsilon^{4})$.

\begin{figure}
\setlength{\unitlength}{0.240900pt}
\ifx\plotpoint\undefined\newsavebox{\plotpoint}\fi
\sbox{\plotpoint}{\rule[-0.200pt]{0.400pt}{0.400pt}}%
\includegraphics[scale=0.5]{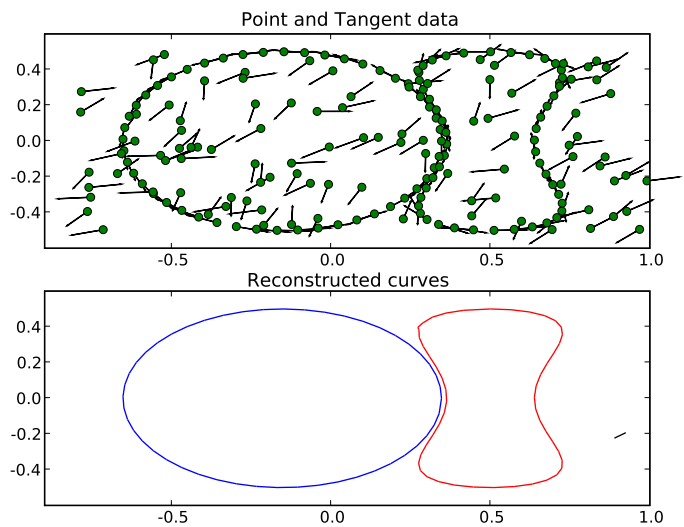}
\caption{The same example as in Figure \ref{fig:basicExample}, but with
100 additional points (for a total of $196$), placed randomly. }
\label{fig:noisyExample}
\end{figure}

\subsubsection{Filtering the data}

The aforementioned criteria suggest that our reconstruction algorithm
has excellent potential for noise removal. It suggests that if we
remove points which do not have edges pointing towards other edges,
then with high probability we are removing spurious edges.

This notion is well supported in practice.
By running Algorithm \ref{algo:polygonalization} on a figure consisting of
$96$ true points, and $100$ randomly placed incorrect points, a nearly
correct polygonalization is calculated (Fig. \ref{fig:noisyExample}).
The original curve is reconstructed with an error at only one point
(the top left corner of the right-hand curve).

Of course, if enough incorrect points are present, some points will
eventually be connected by Algorithm \ref{algo:polygonalization}.
This can be seen in Figure \ref{fig:noisyExample}:
the line segment near $(0.9, -0.2)$ is an edge between two incorrect points.
One hint that an edge is incorrect is that it points to a leaf.
That is, consider a set of vertices $\vp_{0}, \vp_{1}, \ldots, \vp_{n}$
as well as $\vq$. Suppose, after approximately computing the
polygonalization, one finds that the graph contains edges
$e_{0} = (\vp_{0}, \vp_{1}), e_{1} = (\vp_{1}, \vp_{2}), \ldots, e_{n-1} =
(\vp_{n-1}, \vp_{n})$ and $e_{n} = (\vp_{n/2}, \vq)$. The vertex $\vq$ is
a leaf, that is it is reachable by only one edge. A polygonalization
of a set of closed curves should not have leaves, suggesting that the
edge $e_{n}$ is spurious.
Thus filtering leaves is a very reasonable heuristic for noise filtering.

One final problem with noisy data worth mentioning is that sometimes,
an incorrect point will be present that lies within the allowed
region of a legitimate point, and closer to the legitimate point
than the adjacent points along the curve. This will prevent the
correct edge from being added. This can be remedied by adding not
only $\vr_{i}^{\pm}$ at Step 3 of the algorithm, but also points for
which $d_{\vm^{\perp}}(\vp_{i})$ whose distance to $\vp_{i}$ is not
much longer than the distance between $\vp_{i}$ and $\vr_{i}^{\pm}$.
With some luck, this procedure combined with filtering out leaves
will approximately reconstruct the correct figure.

\vspace{.2in}

\hrule
\begin{algo}
  \label{algo:noisyPolygonalization}
  \begin{center}
  {\bf (Polygonalization with Noise Removal)}
  \end{center}

\vspace{.1in}

\hrule

\vspace{.2in}

\noindent { \bf Input: }

[ We assume we are given the dataset $\allData$ (which includes spurious data),
the maximal curvature
$\curvemax$, the noise amplitudes $\pointNoise, \tanNoise$, and a
parameter $\epsilon$ satisfying both $\epsilon \curvemax < 1/\sqrt{2}$ and
$2 \curvemax \epsilon^{2} < \curvesep$.
We assume that adjacent points on a given curve
are less than a distance $\epsilon$ apart, i.e. the curve is
$\epsilon$-sampled. We also assume we are given the number of
leaf removal sweeps $l \in \mathbb{Z}^{+}$ and a
threshold $\alpha \geq 1$. ]

\vspace{.1in}

  \begin{enumerate}
  \item Compute the graph $G = (\pointData, E)$ with edge set:
    \begin{equation*}
      E = \{ (\vp_{i},\vp_{j}) : \vp_{i} \in \allowed{\epsilon}{\vp_{j}} \textrm{~and~} \vp_{i} \in \allowed{\epsilon}{\vp_{j}}\}
    \end{equation*}
  \item For each vertex $\vp_{i} \in \pointData$:
    \begin{enumerate}[a.]
    \item Compute the set of vertices
      \begin{equation*}
        R^{\pm}_{i} = \{ \vp_{j} : (\vp_{i}, \vp_{j}) \in E \textrm{~and~} \pm (\vp_{j}-\vp_{i}) \cdot \vm_{i} > 0 \}
      \end{equation*}
    \item Find the nearest tangential neighbors, i.e.
      \begin{equation*}
        \vr^{\pm}_{i} = \textrm{argmin}_{\vq \in R^{\pm}_{i}} \pm (\vp_{j}-\vp_{i}) \cdot \vm_{i}
      \end{equation*}
    \item Find the set of almost-nearest tangential neighbors:
      \begin{equation*}
        \mathbf{R}^{\pm}_{i} = \{ \vr \in R^{\pm}_{i} : d_{\vm_{i}}(\vp_{i}, \vr) \leq \alpha \vr^{\pm}_{i} \}
      \end{equation*}
    \end{enumerate}
  \item Compute the graph $\Gamma = ( \pointData, E')$ with
    \begin{equation*}
      E' = \bigcup_{i} \{ (\vp_{i}, \vr) : \vr \in \mathbf{R}^{+}_{i} \} \cup \{ (\vp_{i}, \vr) : \vr \in \mathbf{R}^{-}_{i} \}
    \end{equation*}
  \item Search through $\Gamma$ for leaves, and remove edges pointing to the leaves. Repeat this $l$ times.
  \item Output $\Gamma$.
  \end{enumerate}
\end{algo}

\begin{figure}
\setlength{\unitlength}{0.240900pt}
\ifx\plotpoint\undefined\newsavebox{\plotpoint}\fi
\sbox{\plotpoint}{\rule[-0.200pt]{0.400pt}{0.400pt}}%
\includegraphics[scale=0.5]{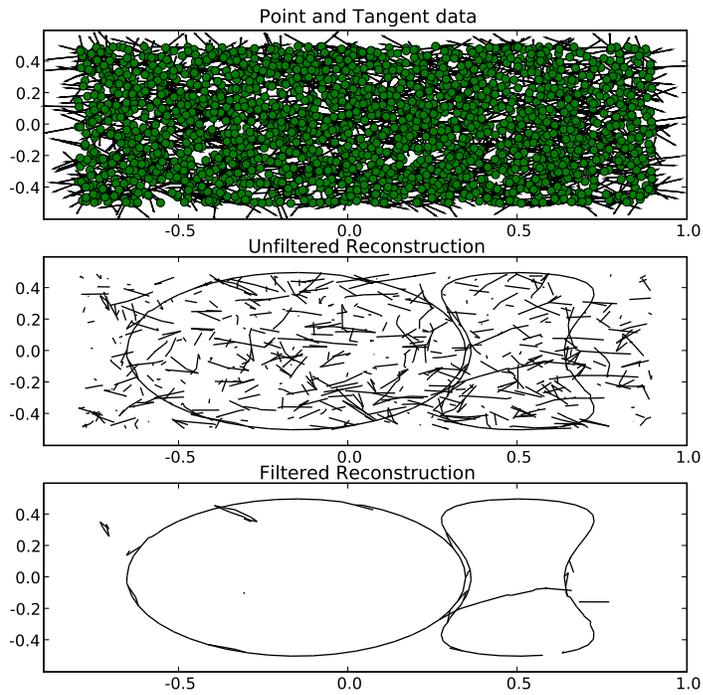}
\caption{The same example as in Figure \ref{fig:basicExample}, but with
2000 additional random points added (for a total of 2096).
The original curve is no longer completely reconstructed, but the
general shape is still roughly visible, along with many more spurious
points. The middle figure shows the reconstruction
without Step 4 of Algorithm 3. Filtering leaves with $l=4$ improves
the situation considerably (bottom figure).
\label{fig:moreNoisyExample}}
\end{figure}

In practice, we have found that $\alpha=1.1$ and
$l=4$ work reasonably well.
Figure \ref{fig:moreNoisyExample} illustrates the result of Algorithm
\ref{algo:noisyPolygonalization}, both with and without filtering.

\section{Conclusions}

Standard methods for reconstructing a finite set of
curves from sample data are quite general.
By and large, they assume that only point samples are given.
In some applications, however, additional information is available.
In this paper, we have shown that if both sample location and
tangent information are given, significant improvements can be made
in accuracy. We were motivated by a problem in medical imaging,
but believe that
the methods developed here will be of use in a variety of other applications,
including MR tractography and contour line reconstruction in topographic
maps \cite{GORE,TOPO}.

\appendix

\section{Proof of Theorem \ref{thm:noisyReconstruction}}
\label{sec:proofOfNoisyReconstruction}
The proof of Theorem \ref{thm:noisyReconstruction} follows that of
Theorem \ref{thm:proofOfAlgo} closely, with minor modifications
made to account for the noise.
To begin, we need to show that the noisy allowed region is large enough
to separate distinct curves. It is here that we use
\eqref{eq:noisySeparationDistance}.

\begin{proposition}
  Suppose $i \neq j$ and assume that
  \eqref{eq:noisySeparationConditions} holds.
  Then $\ball{\pointNoise}{\vp_{i}} \cap
  \nallowed{\epsilon}{\vp_{j}} = \emptyset$ unless $\vp_{i}$ and $\vp_{j}$
  are samples from the same curve, and are separated by an arc length no
  larger than $\curvemaxi \pi/2$.
\end{proposition}

\begin{proof}
  For simplicity, suppose that $\vp_{i} \in \ball{\epsilon}{\vp_{j}}$ (since otherwise, $\vp_{i} \nin \nallowed{\epsilon}{\vp_{j}}$, but $\vp_{i}$ is not sampled from the same curve as $\vp_{j}$. Let $\gamma(t)$ denote the curve from which $\vp_{j}$ is sampled. Let $\vp = \vp_{j}$ and $\vm = \vm_{j}$ to simplify notation.

  Select points $\vp'', \vm''$ to minimize $d(\vp_{i},L'')$, where $L''=\vp''+\vm''$, with the constraint that $d(\vp'',\vp) < \pointNoise$ and the angle between $\vm''$ and $\vm$ is smaller than $\tanNoise$. Let $\va \in L''$ be the point for which $d(\va,\vp_{i}) = d(\vp_{i},L'')$.

  It is shown in the proof of Lemma \ref{lem:separationAllowedRegions} that if $d(\vx, L'') > \curvemax \epsilon^{2}$, then $vx \nin \allowed{\epsilon}{\vp''}$ (recall \eqref{eq:1}, \eqref{eq:4}). Thus, if $d(\vp_{i},L'') > \curvemax \epsilon^{2} + \pointNoise$ for any $\vp'', \vm''$, then $\vx \nin \allowed{\epsilon}{\vp''}$ for each $\vx \in \ball{\pointNoise}{\vp_{i}}$ and hence $\vp_{i} \nin \nallowed{\epsilon}{\vp}$. We will show this to be the case.


  By the second triangle inequality, we have the bound:
  \begin{multline}
    \label{eq:7}
    d(\vp_{i},L'') = d(\vp_{i},\va) \geq d(\vp_{i}',\va) - d(\vp_{i}, \vp_{i}')\\
    \geq d(\vp_{i}', \vb) - d(\vb,\va) - d(\vp_{i}, \vp_{i}')
    \geq \curvesep - d(\vb,\va) - \pointNoise
  \end{multline}
  where $\vb$ is the point on $\gamma(t)$ closest to $\vp_{i}'$. Once we show this is greater than $\curvemax \epsilon^{2}$, the proof is complete.

Let $L'=\vp' + \vm' t$ (with $\vp'$ and $\vm'$ being true samples of $\gamma(t)$, approximated by $\vp$ and $vm$). Then we have the bound:
  \begin{equation}
    \label{eq:6}
    d(\vb,\va) \leq \sup_{\vg \in L'} d(\vb,\vg) + d(\vg, \va) \leq d(\vb,L') + d(\va,L') \leq \curvemax \epsilon^{2} + d(\va,L')
  \end{equation}
  The bound on $d(\vb, \vg)$ follows since $\vb$ is a sample from $\gamma(t)$ (recalling \eqref{eq:1}, \eqref{eq:4}).

  Since $\va=\vp'' + \vm'' s$ (for some $s \in [-\epsilon,\epsilon]$), we can perform the bound:
  \begin{multline}
    \label{eq:5}
    d(\va,L') = \sup_{\abs{t} \leq \epsilon} d(\va, \vp' + \vm' t) \leq \sup_{\abs{s} \leq \epsilon} \sup_{ \abs{t} \leq \epsilon} d(\vp'' + \vm'' s, \vp' + \vm' t) \\
    \leq d(\vp'' + m'' \epsilon, \vp' + \vm' \epsilon) \leq 2 \pointNoise + 4 \tanNoise \epsilon
  \end{multline}
  In \eqref{eq:5}, we assume $m'$ and $m''$ are oriented the same way. It is easy enough to see that the sup is achieved at the endpoints; we then use the triangle inequality $d(\vp'', vp') < d(\vp'', \vp) + d(\vp, \vp') \leq 2 \pointNoise$, and similarly for the tangents.
  Thus, we find that:
  \begin{equation}
    \eqref{eq:6} \leq \curvemax \epsilon^{2} + 2 \pointNoise + 4 \epsilon \tanNoise
  \end{equation}
  Plugging this into \eqref{eq:7} shows that:
  \begin{equation}
    d(\vp_{i},L'') \geq \curvesep - (4 \pointNoise + 4 \epsilon \tanNoise + \curvemax \epsilon^{2}) \geq 1.1 \curvemax \epsilon^{2} > \curvemax \epsilon^{2} + \pointNoise
  \end{equation}
  where the last inequality follows from \eqref{eq:noisySeparationDistance}.
\end{proof}

This shows that the graph $G$ computed in Step 1 separates distinct curves.

The next result parallels Proposition \ref{prop:polyIncludesNeighboringPoints}, and shows that the noisy allowed region contains nearby points on the polygonalization.

\begin{proposition}
  Suppose the figure is sampled at a rate satisfying \eqref{eq:constraintOnkmaxEpsilon}. Then $G$ contains the polygonalization of the figure.
\end{proposition}

\begin{proof}
  The point $\vp_{i}$ and tangent $\vm_{i}$ are close to some point $\vp_{i}', \vm_{i}'$ on the figure $\curveSet$; in particular, $\abs{\vp_{i} - \vp_{i}'} \leq \pointNoise$ and $\arccos(\vm_{i} \cdot \vm_{i}') < \tanNoise$ . Similarly, there is a point $\vp_{j}'$ on the figure a distance no more than $\pointNoise$ away from $\vp_{j}$. By Proposition \ref{prop:polyIncludesNeighboringPoints}, $\vp_{j}' \in \allowed{\epsilon}{\vp_{i}'}$. Since $\vp_{j}' \in \ball{\pointNoise}{\vp_{j}}$ and $\vp_{j}' \in \allowed{\epsilon}{\vp_{i}'} \subset \nallowed{\epsilon}{\vp_{i}}$, we find that $\vp_{j}' \in \ball{\pointNoise}{\vp_{j}} \cap \nallowed{\epsilon}{\vp_{i}} \neq \emptyset$. Repeating this argument with $i$ and $j$ interchanged shows that \eqref{eq:noisyConditionForCheckingIfEdgeConnectionIsPlausible} holds, and $(\vp_{i}, \vp_{j})$ is an edge of $G$.
\end{proof}

\begin{proposition}
  \label{lem:noisyClosestTangentPointInAllowedRegionIsCorrect}
  Fix a point $\vp_{i}=\vp \in \pointData$, and suppose that Assumption \ref{ass:minSamplingRateNoisy} holds. Choose a tangent vector $\vm_{0}$ and fix an orientation. Consider the set of points $\vp_{j}$ such that $(\vp, \vp_{j})$ is an edge in $G$ (as per Step 1 of Algorithm \ref{algo:polygonalizationNoisy}) and $(\vp_{j} - \vp) \cdot \vm_{0} > 0$. Suppose also that $\epsilon$ satisfies \eqref{eq:noisyConstraintOnkmaxEpsilon}.

  Then the nearest tangential neighbor of $\vp$ (i.e. the edge for which $(\vp_{j} - \vp) \cdot \vm_{0}$ is minimal) is the edge in the polygonalization of $\gamma$.
\end{proposition}

\begin{proof}
  The idea of the proof follows that of Lemma \ref{lem:closestTangentPointInAllowedRegionIsCorrect} closely, but we must adjust for our uncertainty as to the point and tangent.

  The curve itself has the parameterization $\gamma(t) = \vp' + \vm' t + \vm'^{\perp} w(t)$, by Lemma \ref{lem:localGraphParameterization}, and this is valid for $\abs{t} < \curvemaxi$. However, we do not know $\vp'$ and $\vm'$, only $\vp$ and $\vm$. We wish to find the point $\vp_{j}$ for which $\vm' \cdot (\vp_{j}' - \vp')$ is minimal, and we approximate this by finding the point for which $\vm \cdot (\vp_{j} - \vp)$ is minimal.

Using the fact that $\vm \cdot (\vp_{j} - \vp) - \vm \cdot (\vp_{k} - \vp) = \vm \cdot (\vp_{j}- \vp_{k})$, we find that
\begin{multline}
  \label{eq:3}
  \vm \cdot (\vp_{j} - \vp) - \vm \cdot (\vp_{k} - \vp) =
  \vm \cdot (\vp_{j}- \vp_{k}) = \\
  \vm \cdot (\vp_{j}' - \vp_{k}') + \vm \cdot ([\vp_{j} - \vp_{j}'] - [\vp_{k} - \vp_{k}']) \\
  = \vm' \cdot (\vp_{j}' - \vp_{k}') + (\vm - \vm') \cdot (\vp_{j}' - \vp_{k}')  + \vm \cdot ([\vp_{j} - \vp_{j}'] - [\vp_{k} - \vp_{k}'])
\end{multline}
The second and third terms on the right side of \eqref{eq:3} are the error terms. We have the bound:
\begin{multline*}
  \abs{(\vm - \vm') \cdot (\vp_{j}' - \vp_{k}')  + \vm \cdot ([\vp_{j} - \vp_{j}'] - [\vp_{k} - \vp_{k}'])} \\
  \leq \sqrt{\sin^{2}(\tanNoise) + (1-\cos(\tanNoise) )^{2}} \abs{\vp_{j}' - \vp_{k}'} + 2 \pointNoise
  \leq 2 (\tanNoise \epsilon + \pointNoise)
\end{multline*}
We wish to find the $j$ for which \eqref{eq:3} is negative for every $k$. If we can show that $\abs{\vm' \cdot (\vp_{j}' - \vp_{k}')} > 2(\tanNoise \epsilon + \pointNoise)$, we are done.

If we observe that $\vp_{j}' = \vp' + \vm' t_{j} + w(t_{j}) \vm'^{\perp}$ (using the notation of Lemma \ref{lem:localGraphParameterization}), and similarly $\vp_{k}' = \vp' + \vm' t_{k} + w(t_{k}) \vm'^{\perp}$, we find then that $\vm' \cdot (\vp_{j}'-\vp_{k}') = t_{j}-t_{k}$.

It is here we use the fact that $\abs{\vp_{j} - \vp_{k}}^{2} = (t_{j}-t_{k})^{2} + (w(t_{j})-w(t_{k}))^{2} \geq [(1+2^{3/2})(2 \tanNoise \epsilon + \pointNoise)]^{2}$. With $f(z)$ as in Lemma \ref{lem:localGraphParameterization}, we find that:
\begin{multline*}
  \abs{(w(t_{j})-w(t_{k}))} = \abs{w'(y)}\abs{(t_{j}-t_{k})} \leq \frac{1}{f'(f^{-1}(\curvemax y))} \abs{(t_{j}-t_{k})} \\
  = (1+(f^{-1}(\curvemax y))^{2})^{3/2} \abs{(t_{j}-t_{k} )}
\end{multline*}
for some $y \in [0,\curvemaxi]$. If $\curvemax y < 1/\sqrt{2}$ (i.e. \eqref{eq:constraintOnkmaxEpsilon} is satisfied), then $f^{-1}(\curvemax y) < 1$ and $(1+f^{-1}(\curvemax y)^{2}) < 2$. Thus:
\begin{multline}
  \abs{(1+2^{3/2})(t_{j}-t_{k})}^{2} \\
  \geq (t_{j}-t_{k})^{2} + [(1+(f^{-1}(\curvemax y))^{2})^{3/2} (t_{j}-t_{k} )]^{2} \\
  \geq \abs{\vp_{j} - \vp_{k}}^{2}  \geq [(1+2^{3/2})(2 \tanNoise \epsilon + \pointNoise)]^{2}
\end{multline}
(the last inequality follows from Assumption \ref{ass:minSamplingRateNoisy}) implying that $\abs{t_{j}-t_{k}} \geq 2(\tanNoise \epsilon + \pointNoise)$.
\end{proof}

\section{Speeding it up: an $O(N \log N)$ algorithm}
\label{sec:quadTreeSection}

As remarked earlier, Algorithm \ref{algo:polygonalization} and \ref{algo:polygonalizationNoisy} run in $O(N^{2})$ time as written. The slow step is Step 1 which involves comparing every point/tangent pair to every other such pair. This scaling issue can be remedied by using a spatially adaptive data structure
\cite{quadtrees}

A caveat: there are two different ways of increasing $N$. The first (increasing outward) is by taking larger figures, with the sampling rate held fixed. The second (increasing inward) is by holding the figure size fixed, but increasing the sampling rate. We are interested primarily in the first case, and we will treat this case only. Therefore, we make the following additional assumption:

\begin{assumption}
  \label{ass:sampleDensity}
  We assume that the density of points in the input data is bounded above, i.e.:
  \begin{equation}
    \sup_{B} \frac{
      \abs{ \pointData \cap B }
    }{
      \abs{B}
    } \leq \densitymax
  \end{equation}
\end{assumption}

Note that this always holds in the case of noisy data. In this case, Assumption \ref{ass:minSamplingRateNoisy} combined with \eqref{eq:noisySeparationConditions} implies that
\begin{equation*}
  \densitymax \leq \frac{1}{ (1+2^{3/2})(2 \tanNoise + \pointNoise) \curvesep} \leq
  \frac{1}{ (1+2^{3/2})(2 \tanNoise + \pointNoise) (4 \pointNoise + 4 \epsilon \tanNoise + 2.1 \curvemax \epsilon^{2})}.
\end{equation*}

In computing Step 1 of Algorithm \ref{algo:polygonalization} or \ref{algo:polygonalizationNoisy}, we must determine whether two points are in each other's allowed region (or a ball of radius $\pointNoise$ about the noisy allowed region). Note that $\allowed{\epsilon}{\vp_{i}} \subset B_{\epsilon}(\vp_{i})$, so if $\abs{\vp_{i} - \vp_{j}} \geq \epsilon$, then clearly the edge $(\vp_{i},\vp_{j}) \nin G$. Similarly, for the noisy case, if $\abs{\vp_{i} - \vp_{j}} \geq \epsilon+2\pointNoise$, then $(\vp_{i},\vp_{j}) \nin G$. We exploit this fact by using quadtrees, which allow us to avoid comparing points more than a distance $\epsilon$ apart.

\vspace{.2in}

\hrule
\begin{algo}
  \label{algo:polygonalization}
  \begin{center} {\bf (Fast Computation of the Graph G)}
  \end{center}

\vspace{.1in}

\hrule

\vspace{.2in}

\noindent { \bf Input: }
[ We assume we are given the dataset $\allData$, the maximal point density $\densitymax$ and the sampling $\epsilon$. We also take the parameter $\lambda=\epsilon$ (noise free case) or $\lambda=\epsilon+2\pointNoise$ (noisy case). ]

\vspace{.1in}

  \begin{enumerate}
  \item Compute a quadtree $Q$ storing $(\vp_{i},\vm_{i})$ pairs. The splitting criteria for a node is when the node contains more than $\densitymax \lambda^{2}$ points.
  \item Initialize the graph $G = (\pointData, E)$ with empty edge set.
  \item For each point $\vp_{i}$, iterate over the points $\vp_{j}$ contained in the node containing $\vp_{i}$ and all of its nearest neighbors. If
    \begin{equation*}
      \vp_{i} \in \allowed{\epsilon}{\vp_{j}} \textrm{~and~} \vp_{j} \in \allowed{\epsilon}{\vp_{i}},
    \end{equation*}
    then add the edge $(\vp_{i},\vp_{j})$ to the graph $G$.
  \item Return $G$.
  \end{enumerate}
\end{algo}
Initializing the quadtree in step 1 is an $O(N \log N)$ operation. Assumption \ref{ass:sampleDensity} implies that the width of a node will be no smaller than $\lambda$; thus, a node containing a point $\vp_{i}$ together with it's nearest neighbors contains the allowed region. The comparison at step 3 involves at most $\densitymax \lambda^{2}$ points, regardless of $N$. Thus, the complexity of this algorithm is
\begin{equation}
  O(N \log(N) + N \densitymax \lambda^{2}).
\end{equation}

\bibliographystyle{hplain}

\begin{thebibliography}{99}

\bibitem{882320}
Bart Adams and Philip Dutr\'{e}.
\newblock Interactive boolean operations on surfel-bounded solids.
\newblock {\em ACM Trans. Graph.}, 22(3):651--656, 2003.

\bibitem{1103907}
Marc Alexa, Markus Gross, Mark Pauly, Hanspeter Pfister, Marc Stamminger, and
  Matthias Zwicker.
\newblock Point-based computer graphics.
\newblock In {\em SIGGRAPH '04: ACM SIGGRAPH 2004 Course Notes}, page~7, New
  York, NY, USA, 2004. ACM.

\bibitem{amenta98crust}
N.~Amenta, M.~Bern, and D.~Eppstein.
\newblock The crust and the $\beta$-skeleton: Combinatorial curve
  reconstruction.
\newblock {\em Graphical models and image processing: GMIP}, 60(2):125, 1998.

\bibitem{amenta98new}
Nina Amenta, Marshall Bern, and Manolis Kamvysselis.
\newblock A new {Voronoi}-based surface reconstruction algorithm.
\newblock {\em Computer Graphics}, 32({Annual Conference Series}):415--421,
  1998.

\bibitem{amenta02simple}
Nina Amenta, Sunghee Choi, Tamal~K. Dey, and N.~Leekha.
\newblock A simple algorithm for homeomorphic surface reconstruction.
\newblock {\em International Journal of Computational Geometry and
  Applications}, 12(1-2):125--141, 2002.

\bibitem{598521}
Rodrigo~L. Carceroni and Kiriakos~N. Kutulakos.
\newblock Multi-view scene capture by surfel sampling: From video streams to
  non-rigid 3d motion, shape and reflectance.
\newblock {\em Int. J. Comput. Vision}, 49(2-3):175--214, 2002.

\bibitem{chengnoise}
Siu-Wing Chen, Stefan Funke, Mordecai Golin, Piyush Kumar,
Sheung-Hung Poon, and Edgar Ramos.
\newblock Curve reconstruction from noisy samples.
\newblock {\em Comput. Geometry}, 31:63--100, 2005.

\bibitem{TOPO}
Y. Chen, R. Wang, and J. Qian.
\newblock Extracting contour lines from common-conditioned topographic maps.
\newblock {\em IEEE Transactions on Geoscience and Remote Sensing},
44(4):1048--1057, 2006.

\bibitem{dey99curve}
Tamal~K. Dey, Kurt Mehlhorn, and Edgar~A. Ramos.
\newblock Curve reconstruction: Connecting dots with good reason.
\newblock In {\em Symposium on Computational Geometry}, pages 197--206, 1999.

\bibitem{dey01reconstructing}
Tamal~K. Dey and Rephael Wenger.
\newblock Reconstructing curves with sharp corners.
\newblock {\em Computational Geometry}, 19(2--3):89--99, 2001.

\bibitem{edelsbrunner}
Herbert Edelsbrunner.
\newblock Shape reconstruction with Delaunay complex.
\newblock In {\em LATIN '98: Theoretical Informatics}, Lecture Notes
in Computer Science, 1380, Springer-Verlag, 119--32, 1998.

\bibitem{quadtrees}
Raphael Finkel and J.L. Bentley.
\newblock Quad Trees: A Data Structure for Retrieval on Composite Keys
\newblock In {\em Acta Informatica}, 4(1): 1--9, 1974.

\bibitem{GORE}
A. Mishra, Y. Lu, A. S. Choe, A. Aldroubi, J. C. Gore,
A, W. Andersona, Z. Ding.
\newblock An image-processing toolset for diffusion tensor tractography.
\newblock {\em Magnetic Resonance Imaging}, 25: 365--376, 2007.

\bibitem{hoppe92surface}
Hugues Hoppe, Tony DeRose, Tom Duchamp, John McDonald, and Werner Stuetzle.
\newblock Surface reconstruction from unorganized points.
\newblock {\em Computer Graphics}, 26(2):71--78, 1992.

\bibitem{1018057}
Thouis~R. Jones, Fredo Durand, and Matthias Zwicker.
\newblock Normal improvement for point rendering.
\newblock {\em IEEE Comput. Graph. Appl.}, 24(4):53--56, 2004.

\bibitem{mdnoise}
Asisn Mukhopadhyay and Augustus Das.
\newblock Curve reconstruction in the presence of noise.
\newblock In {\em CGIV 2007: 4th International Conference on
Computer Graphics, Imaging and Visualization}, pp. 177--182,
IEEE Compuer Society, 2007.

\bibitem{344936}
Hanspeter Pfister, Matthias Zwicker, Jeroen van Baar, and Markus Gross.
\newblock Surfels: surface elements as rendering primitives.
\newblock In {\em SIGGRAPH '00: Proceedings of the 27th annual conference on
  Computer graphics and interactive techniques}, pages 335--342, New York, NY,
  USA, 2000. ACM Press/Addison-Wesley Publishing Co.

\bibitem{383300}
Matthias Zwicker, Hanspeter Pfister, Jeroen van Baar, and Markus Gross.
\newblock Surface splatting.
\newblock In {\em SIGGRAPH '01: Proceedings of the 28th annual conference on
  Computer graphics and interactive techniques}, pages 371--378, New York, NY,
  USA, 2001. ACM.

\end{thebibliography}

\end{document}